\documentclass[conference]{IEEEtran}

\usepackage{amsmath, amssymb} % This package is added to write equations
\usepackage{enumitem} % For customizing list environments
\usepackage{calc} % For measuring the width of the widest label
\usepackage{array} % For increasing the gap between each row/line in the tabular format
\usepackage{cite} % For compressed citation of multiple references
\usepackage{hyperref} % For creating hyperlink to the references
\usepackage{bm} % For bold formatting
\usepackage{booktabs} % For tables
\usepackage{graphicx} % For inserting image
\usepackage{float} % Used for controlling the placement of floating objects, i.e., figures and tables
\usepackage{tabularx} % This package is for tables
\usepackage{multirow}
\usepackage{makecell}
\usepackage{ragged2e}
\allowdisplaybreaks

\begin{document}
\title{Models of Electric Vehicle Charging Demands in Distribution Grid Operation: A Review}
%A review of models of Electric Vehicle Charging Demand in Distribution Grid Operation
\author{\IEEEauthorblockN{Dipayan Sarkar}
\IEEEauthorblockA{Electrical and Computer Engineering\\
University of Central Florida\\
Orlando, USA\\
dipayan.sarkar@ucf.edu}
\and
\IEEEauthorblockN{Qifeng Li}
\IEEEauthorblockA{Electrical and Computer Engineering\\
University of Central Florida\\
Orlando, USA\\
Qifeng.Li@ucf.edu}
}

\maketitle

\begin{abstract}
The rapid integration of electric vehicles (EVs) into the transportation system has introduced a new class of load into power distribution systems (PDS), i.e., EV charging demand (EVCD). The spatiotemporal variability of EVCD poses fundamental challenges for distribution system operators (DSOs), particularly in ensuring secure and reliable network operation. Accurately representing EVCD as a parameter within the distribution optimal power flow (DOPF) problem is a prerequisite for reliable, cost-efficient PDS operation, yet no universally accepted modeling framework has emerged. The existing contributions span a wide spectrum of assumptions, mathematical formulations, and solution paradigms. This article presents a systematic review of methods and models employed to represent EVCD within DOPF frameworks from the DSO's operational perspective. The existing EVCD modeling methods are classified into two primary categories based on EVCD representation: deterministic models and uncertainty-aware models. The two main categories are further divided into several subcategories. This paper aims to provide a comprehensive review on models for integrating EVCD into PDS operation and planning. 
\end{abstract}

%This paper also discusses the solution methods used to solve these models.

%The purpose of this article is to provide a comprehensive review about the methods and models that had been used to model EVCD and how they are handled from the DSO's perspective.

\vspace{-6pt}
\section{Introduction}
\begin{figure}[t]
    \centering
    \includegraphics[width=0.9\linewidth]{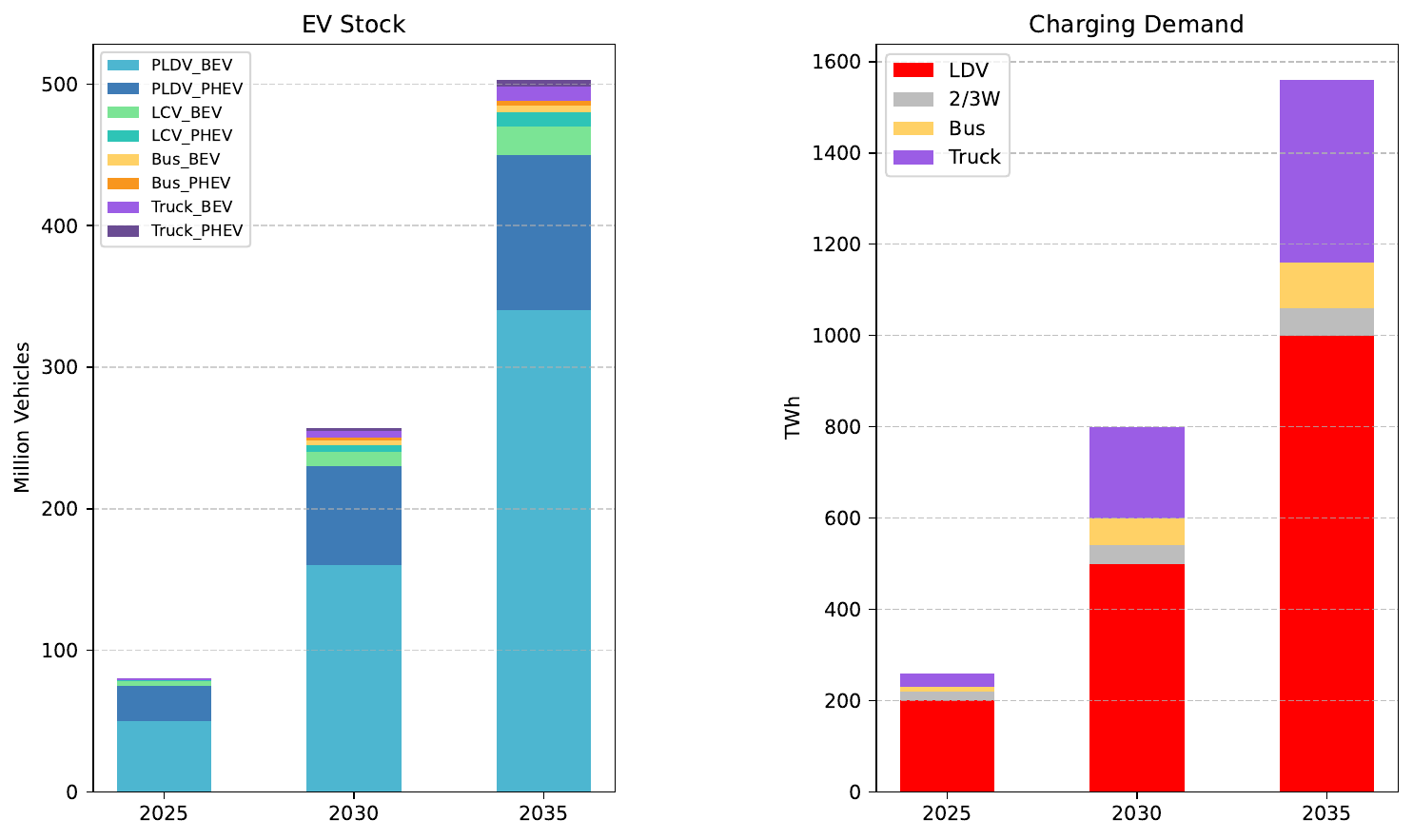}
    \vspace{-10pt}
    \caption{Temporal Evolution of EV Stock and Charging Demand\\
    \tiny
    Notes: BEV = battery electric vehicle; PHEV = plug-in
    hybrid electric vehicle; LDV = light-duty vehicle; PLDV = passenger LDV; LCV = light commercial vehicle; 2/3W = Two/Three Wheeler
    }
    \vspace{-15pt}
    \label{fig:ev_stock_and_CD_projection}
\end{figure}

\vspace{-6pt}
With the global transition toward clean energy, electric vehicle (EV) adoption has accelerated significantly in recent years. Global EV charging demand (EVCD) reached 250~TWh in 2025 and is projected to grow to 800~TWh by 2030, potentially exceeding 1500~TWh by 2035 \cite{IEA}, as illustrated in Fig.~\ref{fig:ev_stock_and_CD_projection}. This trajectory implies that EVCD, which accounted for approximately 1\% of global electricity consumption in 2025, will increase nearly sixfold within a decade and will place a substantial stress on power distribution system (PDS).

EVCD can be modeled based on driver mobility patterns \cite{kang2022mapping}, arrival time (AT) and departure time (DT) \cite{zhang2024bilevel}, initial state-of-charge (SOC) of the battery \cite{ren2024two}, and the required energy \cite{jian2023analytical}. Researchers proposed different methods to model EVCD in some of the existing literature that are either from EV aggregator (EVA) perspective \cite{sharma2021risk} or for EV charging station (EVCS) planning \cite{Jin2024} without showing how to integrate them into the PDS operation \cite{li2021uncertainty}. However, if EVCD is not properly modeled and incorporated into the PDS operation, PDS will experience voltage violations, thermal overloading of lines and transformers, increased power losses, and degraded power quality \cite{tayri2025grid, Nutkani2024}. Therefore, this review focuses on models that explicitly formulate EVCD and integrate it into distribution optimal power flow (DOPF) models. 

To ensure that this review reflects recent developments in EV charging demand modeling, the literature was collected from Scopus, Web of Science, ScienceDirect, IEEE Xplore, and Google Scholar using combinations of terms including EV charging demand, uncertain EV demand, machine learning-based EV charging, EV demand uncertainty, EV demand in distribution systems, and EV charging loads. The search was conducted across titles, abstracts, and keywords of the databases, and only publications from 2018 onward were included.

The remainder of this paper is organized as follows. Section~II presents an overview of the reviewed EVCD models and the motivation behind their classification. Section~III details the EVCD models across all categories. Section~IV discusses research gaps and future directions before making concluding remarks in Section~V.  

\vspace{-3pt}
\section{EVCD Modeling: Overview and Classification}
\vspace{-3pt}
\begin{table}[!t]
\renewcommand{\arraystretch}{1.45}
\centering
\caption{EVCD Model Classification Based on Uncertain EV Parameters}
\label{tab:evcd_models}
\scriptsize
\setlength{\tabcolsep}{3pt}
\begin{tabular}{%
  >{\centering\arraybackslash}p{1.35cm}   % Ref
  >{\RaggedRight\arraybackslash}p{1.85cm} % Uncertain Parameter
  >{\RaggedRight\arraybackslash}p{1.9cm}  % DOPF Formulation
  >{\centering\arraybackslash}p{1.5cm}    % Uncertainty Type
}
\toprule
\textbf{Ref.} &
\textbf{Uncertain Parameter} &
\makecell{\textbf{DOPF}\\\textbf{Formulation}} &
\makecell{\textbf{Uncertainty}\\\textbf{Type}} \\
\midrule
 
%% ---- Arrival Time & Departure Time ----
\cite{Giraldo2023}, \cite{Rawat2024},
  \cite{Shi2023}, \cite{li2022temporal} &
Arrival Time \& Departure Time &
SO, RO, DRO &
ExU \\
\midrule
 
%% ---- Daily Mileage ----
\cite{Shi2020}, \cite{Liu2018}, \cite{Wang2025robust}, \cite{quijano2021increasing} &
Travel distance &
SO, CCO, DRO &
Both ExU \& EnU \\
\midrule
 
%% ---- EV Fleet Energy Bounds ----
\cite{Feizi2022}, \cite{zhu2025day}, \cite{Silva2024} &
EV Fleet Energy Bounds &
CCO, RO &
ExU \\
\midrule
 
%% ---- Charging Energy Demand ----
\cite{Jin2024}, \cite{Liu2025}, \cite{Nguyen2023}, \cite{Li2025}, \cite{Zhao2024} &
Charging Energy Demand &
SO, CCO, RO, DRO &
Both ExU \& EnU \\
\midrule
 
%% ---- Traffic Flow / Spatial Demand ----
\cite{wang2023stochastic}, \cite{Xiang2024}, \cite{kong2022spatial}&
Traffic Flow / Spatial Demand &
SO, DRO &
ExU \\
\midrule
 
%% ---- EV Owner Behavioral Parameters ----
\cite{Wang2025}, \cite{zhang2024joint}, \cite{Wang2025V2G}, \cite{sarkar2026promethee} &
EV Owner Behavioral Param. &
CCO, DRO, DRCCO &
Both ExU \& EnU \\
 
\bottomrule
\end{tabular}
\vspace{-15pt}
\end{table}

The accurate representation of EVCD is essential for any credible DOPF formulation from the distribution system operator (DSO) perspective \cite{tirunagari2022reaping}. However, no generalized modeling framework currently exists that can capture EVCD with complete accuracy and certainty. As a result, EVCD modeling within DOPF spans a broad spectrum, ranging from deterministic representations \cite{Sharma2024} to sophisticated uncertainty-aware formulations that capture the stochastic and decision-dependent nature of EV charging behavior \cite{sarkar2026promethee}.

Deterministic EVCD models represent the most tractable class due to their computational simplicity, but they often fail to accurately reflect real-world variability in EVCD \cite{wang2025stochastic}. To improve prediction accuracy, machine learning (ML)-based approaches have been proposed, where historical data are utilized to forecast future charging demand \cite{Alaraj2025, Shokati2024}. In addition, several deterministic formulations incorporate specific EV characteristics, enabling EVCD to be modeled either by the EVA or the DSO, either directly \cite{shao2022generalized} or through coordination with the transportation system operator (TSO) \cite{qiao2022distributed}. The resulting EVCD is subsequently treated as a known input parameter in the DOPF problem. 

\begin{figure}[!t]
    \centering
    \includegraphics[width=1\linewidth]{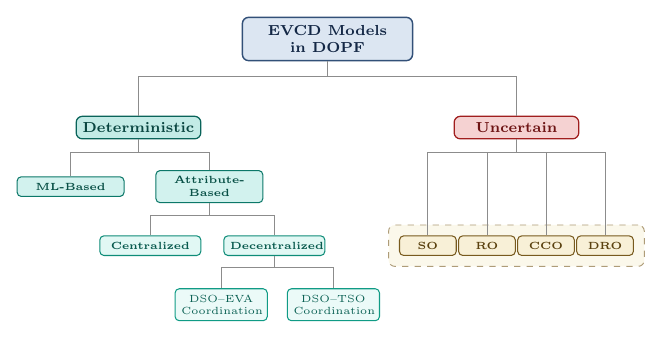}
    \vspace{-25pt}
    \caption{Taxonomy of EVCD Models in DOPF}
    \vspace{-5pt}
    \label{fig:tax_evcd_dopf}
    \vspace{-10pt}
\end{figure}

However, in practical systems, complete information about EV-related parameters is not available to any system operator. Uncertainty arises from multiple sources, including, but not limited to AT, DT and SOC. Such uncertainties may originate from external sources, i.e., exogenous uncertainty (ExU) \cite{Giraldo2023, Rawat2024, Shi2020, Shi2023} or be induced by system-level decisions, i.e., endogenous uncertainty (EnU) \cite{Wang2025, Wang2025V2G, Wang2025robust, sarkar2026promethee}. These uncertainties in EVCD are incorporated into DOPF formulations and addressed using established optimization paradigms, including stochastic optimization (SO), robust optimization (RO), chance-constrained optimization (CCO), distributionally robust optimization (DRO), and distributionally robust chance-constrained optimization (DRCCO). A taxonomy of EVCD modeling approaches is illustrated in Fig. \ref{fig:tax_evcd_dopf}. Furthermore, a detailed classification of EVCD models with respect to the uncertain parameters and associated type of uncertainty is summarized in Table \ref{tab:evcd_models}.

\section{Mathematical Models Used in DOPF}
Accurate modeling of EVCD within the DOPF framework is mandatory for DSO for proper operational decision-making. Given the spatiotemporal variability of EV charging loads, an inadequate representation of EVCD may lead to suboptimal dispatch decisions and potentially compromise the secure operation of PDS. In this review, emphasis is placed on studies that explicitly integrate EVCD models into DOPF formulations. The principal modeling approaches are outlined as follows.
\vspace{-6pt}
\subsection{Deterministic Models} 
\vspace{-3pt}
\subsubsection{Attribute-Based EVCD models}
In addition to data-driven approaches, EVCD can be modeled using intrinsic characteristics of EV, such as AT, DT, SOC, and charging price. Depending on the system architecture, EVCD may be determined either directly by DSO or by EVA or by TSO, with possible coordination mechanisms among them.

\paragraph{Centralized Optimization}
In this framework, the DSO has access to all the information required for determining EVCD and does not rely on coordination with external entities. The EVCD is modeled based on either unidirectional charging behavior \cite{Zheng2019, das2020charging, shao2022generalized} or bidirectional charging and discharging operations \cite{Tian2024, Ding2020}. The EVCD model is jointly formulated with the DOPF problem as a unified optimization framework. For brevity, the DOPF formulation is omitted, and only the EVCD model is presented as follows:
\vspace{-6pt}
\begin{IEEEeqnarray}{l}
    P_{t}^{\text{load, agg}}=P_{t}^{\text{load}}-P_{t}^{Cdis}+P_{t}^{Ccha} \IEEEyesnumber \IEEEyessubnumber* \label{eq:cental_a} \\
    P_{t}^{Cdis}=\sum_{w \in W^{ch}} P_{w, t}^{\text{dis}} K_{w,t}, \quad t \in T^{c h} \label{eq:cental_b} \\
    P_{t}^{Ccha}=\sum_{w \in W^{ch}} P_{w, t}^{\text{cha}} K_{w,t}, \quad t \in T^{ch} \label{eq:cental_c} \\
    \mathrm{SOC}_{w, t}=\mathrm{SOC}_{w, t-1} +\left(P_{w, t}^{\text{cha}} \eta_{\text{cha}}-P_{w, t}^{\text{dis}} / \eta_{\text{dis}}\right) / E_{\text{cap}, w} \IEEEeqnarraynumspace \IEEEyessubnumber*  \label{eq:cental_e} \\
    \mathrm{SOC}_{w, T_{\text{out}, w}} \geq \mathrm{SOC}_w^{\text{exp}}, \quad w \in W^{c h} \label{eq:cental_g} \\
    \sum_{w \in W^{ch}} K_{w, t} H_{w,t} \leq S, \quad t \in T^{ch} \label{eq:cental_j} 
    \vspace{-5pt}
\end{IEEEeqnarray}
where $P_{t}^{Ccha}, P_{t}^{Cdis}, P_{t}^{\text{load}}, P_{t}^{\text{load, agg}}$ are the charging and discharging load of EV, normal load, and the aggregated load, respectively. The parameter $K_{w, t}$ indicates if an EV $w$ is plugged in EVCS at time $t$, $H_{w,t}$ indicates the charging/discharging state at time $t$ and $E_{\text{cap}, w}$ is the battery capacity. $\mathrm{SOC}_w^{\text{exp}}, \mathrm{SOC}_{w, T_{\text{out}, w}}, \mathrm{SOC}_{\text{min}}, \mathrm{SOC}_{\text{max}}$ are the expected level, final SOC level while leaving the EVCS and the minimum and maximum SOC levels, respectively. The sets of EVs and time periods are indicated by $W^{ch}, T^{ch}$ respectively. Constraint \eqref{eq:cental_a} indicates the aggregated load with the normal load and EVs; constraints \eqref{eq:cental_b} and \eqref{eq:cental_c} give the integrated discharging and charging powers in the EVCS. Equation \eqref{eq:cental_e} describes the energy storage process corresponding to the charging and discharging powers for the $w$th EV; \eqref{eq:cental_g} indicates that the SOC for the $w$th EV should achieve a certain level once the EV leaves at $T_{out,w}$. Equation \eqref{eq:cental_j} guarantees that the capacity of the EV load demand should be limited by the maximum allowable power $S$ in the station.

\paragraph{DSO and EVA Coordination Model}
In a decentralized optimization framework, DSO does not possess complete information regarding EV characteristics, such as AT, DT, and SOC. Instead, this information is available to the EVA, which manages the EV fleet. Consequently, the DSO coordinates with the EVA to obtain the aggregated EVCD at each time period \cite{Wu2026, Xiao2024, Wu2022, huang2025facilitating, suri2026eva, ji2025tri}. The EVA pursues its own profit-maximization objective by submitting energy bids for the managed EV fleet over the scheduling horizon and coordinating slow and fast charging operations to reduce peak demand. The overall process involves two interconnected optimization problems—one for the DSO and another for the EVA. EVA’s optimization model is formulated as follows:
\vspace{-5pt}
\begin{IEEEeqnarray}{l}
    \max_{P_t} U_{eva} =\sum_{T} (\lambda_t P_t-\mu\left(P_t-p_t\right)^2 ) \IEEEyesnumber \IEEEyessubnumber* \label{eq:decentral_a} \\
    \text { s.t. } \quad P_t=\sum_{w=1}^{W^{ch}} x_{w, t} \IEEEyessubnumber* \label{eq:decentral_b} \\ 
    \lambda_t = y_t P_t + z_t \label{eq:decentral_c}
    \vspace{-5pt}
\end{IEEEeqnarray}
where $x_{w, t}$ is the required charging energy of an EV $w$, $\lambda_t$ is the charging price, $p_t$ is the load target and $P_t$ is the total charging demand at time $t$. A penalty parameter $\mu$ is used for deviation from the load target and $y_t$ and $z_t$ indicates the slope and intercept of the pricing function. The charging energy of an EV owner $w$ is determined from EV owner's optimization problem. 
\vspace{-8pt}
\begin{IEEEeqnarray}{l}
    \max_{x_{w, t}}  \sum_{T} U_{w,t} = \sum_{T} (-\frac{1}{2} \beta_{w,1} x_{w, t}^2 + \beta_{w,2} x_{w, t} - \lambda_t x_{w, t}) \IEEEyessubnumber* \IEEEeqnarraynumspace \label{eq:decentral_d} \\
    \text { s.t. } \quad \sum_{t \in T_{GC}} x_{w, t}=E_w \label{eq:decentral_e} 
    \vspace{-5pt}
\end{IEEEeqnarray}
where $U_{w,t}$ is the utility of an EV owner, $E_w$ is the charging requirement, and $T_{GC}$ is grid-connection window.

\paragraph{DSO and TSO Coordination Model}
In this approach, the DSO and the TSO independently optimize the PDS and the transportation system (TS), respectively, while exchanging information through a defined coordination interface. The coupling between the two systems is captured via a boundary condition, which represents the EVCD generated from the TS owing to traffic flow dynamics \cite{qiao2022distributed, Sharma2024, Yousuf2024, Yuan2026, sadhu2022optimal}. This boundary condition serves as the information exchange variable between the two optimization problems, enabling the DSO to incorporate traffic-driven charging demand into the DOPF formulation without requiring direct access to transportation system information. The optimization problems of the TSO and DSO are formulated as follows:
\vspace{-6pt}
\begin{IEEEeqnarray}{l}
    \text{(TSO)} \min _{\boldsymbol{x}_{\mathbf{v}} \in \mathcal{X}_{\mathrm{v}}, \alpha_{\mathbf{v}} \in\{0,1\}}  f_{\mathrm{v}}\left(x_{\mathbf{v}}\right) :=\sum r C \IEEEyesnumber \IEEEyessubnumber* \IEEEeqnarraynumspace \label{eq:TS&DS_a} \\
    \text{s.t.} \quad P^T=z \label{eq:TS&DS_b} 
    \vspace{-6pt}
\end{IEEEeqnarray}
where $\boldsymbol{x}_{\mathbf{v}}, \alpha_{\mathbf{v}}$ represents the set of all decision variables and binary variables, $\mathcal{X}_{\mathrm{v}}$ indicates the set of all constraints of TS, $r$ indicates traffic flow and $C$ is the travel cost. The EVCD is denoted by $P^T$ and is derived using the following equation:
\vspace{-6pt}
\begin{IEEEeqnarray}{l}
    P_{EV} = z = \sum r E_{EV}^{ch}, \IEEEyessubnumber* \label{eq:TS&DS_c} 
    \vspace{-6pt}
\end{IEEEeqnarray}
where $P_{EV}$ is the total EVCD and $E_{EV}^{ch}$ is the required charging energy of an EV. This EVCD is used as a deterministic parameter in the DSO optimization model. DSO optimization model is as follows:
\begin{IEEEeqnarray}{l}
    \text{(DSO)} \min _{\boldsymbol{x}_{\mathbf{p}} \in \mathcal{X}_{\mathrm{p}}, \alpha_{\mathbf{p}} \in\{0,1\}}  f_{\mathrm{p}}\left(x_{\mathrm{p}}\right):=\sum c P \IEEEyessubnumber*  \label{eq:TS&DS_d} \\
    \text{s.t.} P^D=z \label{eq:TS&DS_e} 
\end{IEEEeqnarray}
where where $\boldsymbol{x}_{\mathbf{p}}, \alpha_{\mathbf{p}}$ represents the set of all decision variables and binary variables in DOPF and $\mathcal{X}_{\mathrm{p}}$ indicates the set of all constraints in PDS. The EVCD is denoted by $P^D$.

\subsubsection{ML-Based Predictive Models}
ML-based predictive models forecast future EVCD from historical load data, and this EVCD is used as a known parameter in the DOPF. ML model forecasting is a two-phase procedure: a training phase in which model parameters are estimated, followed by an inference phase in which demand forecasts are generated over the dispatch horizon. Supervised methods such as extreme gradient boosting, linear regression \cite{Frendo2021}, support vector machine, and random forest \cite{zhang2026charging} have been widely employed for short-term forecasting, while unsupervised approaches including autoregressive integrated moving average \cite{Amini2016} and K-means clustering \cite{xiong2018electric} have been applied to demand characterization and load profile extraction. Deep learning architectures including long short-term memory networks \cite{Wang2023, Yi2022}, deep generative models \cite{Wang2020}, artificial neural networks \cite{singh2026energizing}, convolutional neural networks \cite{Zhang2021}, transformer models \cite{hussain2025short}, ensemble methods \cite{shahriar2021prediction}, and transfer learning approaches \cite{roozkhosh2026bridging}, have been proposed for spatiotemporal EVCD forecasting, owing to their capacity to capture nonlinear temporal dependencies.  

\subsection{Uncertainty-aware Models}
In practical systems, EVCD is inherently uncertain due to variability in factors such as electricity price, SOC, AT, DT, EV owner preferences and decisions. To capture these uncertainties, a range of modeling approaches has been proposed in the literature. In this review, uncertainty-aware EVCD models are classified into four categories based on the underlying solution methodology.

\subsubsection{SO Model}
In SO-based EVCD models, key EV parameters are treated as random variables characterized by known probability distributions. Parameters commonly modeled as stochastic include AT, DT, required charging energy, and initial SOC, which are represented using distributions such as Poisson \cite{amara2023modelling, Giraldo2023}, uniform \cite{Giraldo2023}, normal \cite{yang2025spatiotemporal}, lognormal \cite{quijano2021increasing}, or truncated normal \cite{Rawat2024, wang2023stochastic}. 
\begin{IEEEeqnarray}{l}
    a_{h, s} \sim \operatorname{Pois} (\eta_h^{\mathrm{a}}) \qquad d_{h, s} \sim \operatorname{Weib} (k, \eta_h^{\mathrm{d}})+a_{h, s} \IEEEyesnumber \IEEEyessubnumber* \label{eq:so_a} \\
    E_{h, s} \sim \mathcal{U}\left(e_1, e_2\right) \label{eq:so_c}
\end{IEEEeqnarray}
where $a_{h, s}, d_{h, s}$ indicate AT and DT respectively and $E_{h, s}$ denote required charging energy.  Monte Carlo simulation method is used to generate scenarios based on this known probability distribution, and DSO formulates the problem as a two-stage stochastic problem.

\subsubsection{RO Model}
In these models, either the initial energy of EV, SOC or EVCD is considered uncertain and an uncertainty set is declared. 
\vspace{-5pt}
\begin{IEEEeqnarray}{l}
    E^{\mathrm{ini}} \equiv \left\{E_w^{\mathrm{ini}}: \sum_{w \in W^{ch}} \frac{\left|E_w^{\mathrm{ini}}-\bar{E}_w^{\mathrm{ini}}\right|}{\hat{E}_w^{\mathrm{ini}}} \leq \Gamma_w ;\right. \notag \\
    \left.E_w^{\mathrm{ini}} \in\left[\bar{E}_w^{\mathrm{ini}}-\hat{E}_w^{\mathrm{ini}}, \bar{E}_w^{\mathrm{ini}}+\hat{E}_w^{\mathrm{ini}}\right] ; \quad \forall w\right\} \IEEEyesnumber \IEEEyessubnumber* \label{eq:ro_a}
\end{IEEEeqnarray}
where $E_w^{\mathrm{ini}}$ is the initial charge of an EV $w$, $\bar{E}_w^{\mathrm{ini}}$ is the average value, $\hat{E}_w^{\mathrm{ini}}$ is the standard deviation, and $\Gamma_w$ indicates \textit{budget of uncertainty}. The average and standard deviation are obtained from historical data. The EVCD is determined by the following equations:
\vspace{-3pt}
\begin{IEEEeqnarray}{l}
    E_{w, 1}^{\mathrm{EV}}=E_w^{\mathrm{ini}}+P_{w, 1}^{\mathrm{EV}} \Delta_t ; \quad \forall w \IEEEyessubnumber* \label{eq:ro_b} \\
    E_{w, t}^{\mathrm{EV}}=E_{w, t-1}^{\mathrm{EV}} + P_{w, t}^{\mathrm{EV}} \Delta_t ; \quad \forall w, t \geq 1 \label{eq:ro_c} \\
    P_{t}^{\mathrm{EV}} = \sum_{\forall w} P_{w, t}^{\mathrm{EV}} \label{eq:ro_d}
    \vspace{-5pt}
\end{IEEEeqnarray}
where $P_{t}^{\mathrm{EV}}$ is the total EVCD for all EV at time $t$ and is used in the power balance equation in DOPF. These models can be found in \cite{Silva2024, Shi2020, Liu2025, Wang2025robust, ren2024two}.

\begin{table}[!t]
\vspace{-7pt}
\caption{Comparison of EVCD Modeling Approaches}
\vspace{-5pt}
\label{tab:comparison}
\centering
\scriptsize
\renewcommand{\arraystretch}{1}
\setlength{\tabcolsep}{6pt}
\begin{tabular}{lcccccc}
\hline
 &
\textbf{Attribute} &
\textbf{ML} &
\textbf{SO} &
\textbf{RO} &
\textbf{CCO} &
\textbf{DRO} \\
\hline
Real-Time        & \checkmark & \checkmark & $\times$ & $\times$ & \checkmark & $\times$ \\
Planning       & $\times$ & $\times$ & \checkmark & \checkmark & \checkmark & \checkmark \\
Comp.       & Low & Low & High & Medium & Medium & High \\
Scal.       & High & High & Medium & High & Medium & Medium \\
\hline
\end{tabular}

\vspace{1mm}
\tiny
\textit{Abbreviations:} Comp. = Computational complexity; Scal. = Scalability;
\vspace{-18pt}
\end{table}

\subsubsection{CCO Model}
In CCO-based EVCD models, uncertain parameters such as driving distance, charging availability, aggregate power injection, and EV owner response are incorporated into the DOPF formulation through probabilistic constraints, which enforce operational feasibility up to a specified confidence level. These chance constraints are rendered tractable via deterministic reformulation \cite{zhu2025day, zhang2024bilevel, jian2023analytical, kong2022spatial, li2022temporal}, conditional value-at-risk (CVaR) approximation \cite{zhang2024joint}, or binary variable representations \cite{Liu2018}. 
\begin{IEEEeqnarray}{l}
    \operatorname{Pr}\left\{\begin{array}{l}
    e_j^{-} \leq \sum_{t^{\prime}=t_0}^t\left(x_{w, t^{\prime}}-d_{w, t^{\prime}}\right)+e_{w, 0} \notag \\
    \sum_{t^{\prime}=t_0}^t\left(x_{w, t^{\prime}}-d_{w, t^{\prime}}\right)+e_{w, 0} \leq e_j^{+} \notag \\
    x_{w, t} \leq p_{+} s_{w, t}
    \end{array}, \forall t \in T \right\}  \geq 1 - \epsilon \notag \\ 
    \forall w \in W^{ch} \IEEEyesnumber \IEEEyessubnumber* \IEEEeqnarraynumspace \label{eq:cco_a}
\end{IEEEeqnarray}
where $e_{w,0}, e_j^{-}, e_j^{+}$ indicate the initial, lower limit and upper limit of EV SOC level respectively. $x_{w, t^{\prime}}, d_{w, t^{\prime}}$ is the charging energy and driving energy consumption respectively of the EV $w$ at time period $t^{\prime}$. The parameter $p_{+}$ indicates upper limit of charging and $s_{w, t}$ is the charging availability indicator of the EV $w$ at time $t$. $S_c$ is the scenario set and $W^{ch}$ is the set of EV owners. A binary variable $z_{w,s}$ is introduced, and the above equation is reformulated as follows:
\vspace{-7pt}
\begin{IEEEeqnarray}{l}
    e_j^{-} \leq \sum_{t^{\prime}=t_0}^t (x_{w, t^{\prime}}-d_{w, s, t^{\prime}} )+e_{w, 0}+z_{w, s} \sum_{t^{\prime}=t_0}^t d_{w, s, t^{\prime}}  \IEEEyessubnumber* \label{eq:cco_b} \\
    \sum_{t^{\prime}=t_0}^t (x_{w, t^{\prime}}-d_{w, s, t^{\prime}} )+e_{w, 0} \leq e_j^{+}+z_{w, s} \sum_{t^{\prime}=t_0}^t (p_{+}-d_{w, s, t^{\prime}}) \notag \\
    \forall w \in W^{ch}, \forall s \in S_c, \forall t \in T \label{eq:cco_c} \\
    x_{w, t} \leq p_{+} s_{w, s, t} + p_{+} z_{w, s} \quad \forall w \in W^{ch}, \forall s \in S_c, \forall t \in T \label{eq:cco_d} \\
    \sum_{s \in S_c} (\pi_{w, s} z_{w, s}) \leq \epsilon, \quad z_{w, s} \in\{0,1\}, \forall w \in W^{ch}  \label{eq:cco_e}
    \vspace{-5pt}
\end{IEEEeqnarray}

\subsubsection{DRO Model}
Both DRO and DRCCO formulations are considered within this class. Rather than assuming a fully known probability distribution, these approaches construct an ambiguity set over which the worst-case expectation or chance constraint is optimized. The ambiguity set is commonly characterized using moment information \cite{sarkar2026promethee, Wang2025, Wang2025V2G, Shi2023}, norm-based divergence measures such as the 1-norm and inf-norm \cite{Zhao2024, Xiang2024}, Wasserstein distance \cite{Nguyen2023, Feizi2022, cao2021robust}, or Kullback–Leibler (KL) divergence \cite{Li2025}. A representative moment-based ambiguity set is formulated as follows:
\vspace{-5pt}
\begin{IEEEeqnarray}{l}
\mathcal{M}_{1}  = 
\Bigg\{ \mathbb{P}\in\mathcal{P}\big(\mathbb{R}^{|N_{evcs}|}\big) : 
\nonumber \\[2pt]
\big(\mathbb{E}_{\mathbb{P}}[\mathbf{p}_s^{ev}] -\bar{\mathbf{p}}^{ev}\big)^{T}
\boldsymbol\Sigma^{-1}
\big(\mathbb{E}_{\mathbb{P}}[\mathbf{p}_s^{ev}] - \bar{\mathbf{p}}^{ev}\big)
\le \gamma_{1}, \nonumber \\
\mathbb{E}_{\mathbb{P}}\Big[
\big(\mathbf{p}_s^{ev}-\bar{\mathbf{p}}^{ev}\big)
\big(\mathbf{p}_s^{ev}-\bar{\mathbf{p}}^{ev}\big)^{T}
\Big]
\preceq \gamma_{2}\,\boldsymbol\Sigma
\Bigg\}
\IEEEyesnumber \IEEEyessubnumber* \label{eq:drcco_a}
\vspace{-5pt}
\end{IEEEeqnarray}
where $\mathcal{P}(\mathbb{R}^{|N_{evcs}|})$ represents the set of all probability distributions on $\mathbb{R}^{|N_{evcs}|}$ and $N_{evcs}$ is total number of EVCS. The parameter $\mathbf{p}_s^{ev}$ indicate the charging demand in scenario s, $\bar{\mathbf{p}}^{ev}$ and $\boldsymbol\Sigma$ denote the empirical mean and covariance of EVCD. Parameters $\gamma_1 \ge 0$ and $\gamma_2 \ge 1$ reflect the confidence level in empirical estimates. The following equation indicates the probabilistic chance constraint.  
\begin{IEEEeqnarray}{l}
    \mathbb{P} (\mathbf{\mathcal{Y}}^{min} \leq \tilde{\mathbf{\mathcal{Y}}} \leq \mathbf{\mathcal{Y}}^{max}) \geq 1-\epsilon \IEEEyessubnumber* \label{eq:drcco_b} 
\end{IEEEeqnarray}
where $\tilde{\mathbf{\mathcal{Y}}}$ is the set of uncertain variables, $\mathbf{\mathcal{Y}}^{min}, \mathbf{\mathcal{Y}}^{max}$ are the lower and upper limits respectively, and $\epsilon$ is the violation level. The above equation is reformulated into deterministic and tractable equations using the ambiguity set.
\begin{IEEEeqnarray}{l}
    \mathbf{\mathcal{Y}} + \eta(\epsilon)(var(\tilde{\mathbf{\mathcal{Y}}})^{1/2} \leq \mathbf{\mathcal{Y}}^{max} \IEEEyessubnumber* \label{eq:drcco_c}  \\
    \mathbf{\mathcal{Y}} - \eta(\epsilon) (var(\tilde{\mathbf{\mathcal{Y}}}))^{1/2} \geq \mathbf{\mathcal{Y}}^{min} \label{eq:drcco_d} 
\end{IEEEeqnarray}
where $var(\tilde{\mathbf{\mathcal{Y}}})$ is the variance of $\tilde{\mathbf{\mathcal{Y}}}$ and $\eta(\epsilon)$ is worst-case quantile that is calculated using $\gamma_1, \gamma_2$ and $\epsilon$. 

\vspace{-5pt}
\section{Research Gap and Discussion}
\vspace{-3pt}
The reviewed literature reveals several persistent gaps in EVCD-DOPF frameworks. A comparison of EVCD modeling approaches is given in Table \ref{tab:comparison}. The overwhelming majority of uncertainty-aware formulations treat EVCD as an exogenous uncertain parameter independent of DSO decisions \cite{Giraldo2023, Rawat2024, Shi2020, Zhao2024}, despite the fact that DSO pricing and dispatch signals directly influence EV owner charging behavior. Although recent contributions have begun addressing this by modeling EVCD as endogenous uncertainty shaped by real-time DSO decisions \cite{Wang2025, Wang2025V2G, Wang2025robust, sarkar2026promethee}, such formulations remain scarce. Furthermore, existing models predominantly assume a homogeneous EV owner population; while isolated efforts have incorporated owner behavior through survey data \cite{yao2026modeling} or willingness-to-charge formulations \cite{hu2019modeling}, explicit integration of behavioral heterogeneity into a coupled EVCD-DOPF framework remains largely absent. A further structural gap exists in the treatment of transportation-power system coupling: studies accounting for TS-PDS interactions have adopted deterministic EVCD representations \cite{sadhu2022optimal}, while uncertainty-aware formulations have largely omitted TS dynamics \cite{li2022temporal, Wang2025robust}. Finally, the diversity of approaches reviewed reflects the absence of a generalized EVCD formulation encompassing all relevant EV parameters and system interactions. Future research should therefore prioritize the development of uncertainty-aware EVCD models that incorporate behavioral heterogeneity and explicitly capture TS-PDS coupling within a unified DOPF framework.

\vspace{-5pt}
\section{Conclusion}
\vspace{-5pt}
This paper has presented a systematic review of EVCD modeling and its integration into the DOPF framework, examined exclusively from the DSO's perspective. The reviewed literature was classified according to the nature of EVCD representation and the solution methodology employed. Accurate EVCD modeling remains a fundamentally challenging problem, as charging demand is governed by multiple interdependent parameters across different systems and cannot be reduced to a simple deterministic expression. Moreover, no consensus has emerged on whether EVCD should be treated as a deterministic parameter, an exogenous uncertainty, or an endogenous uncertainty, and no existing model comprehensively accounts for all relevant parameters and system interactions. By systematically surveying and categorizing the existing methods and their solution approaches from the DSO's perspective, this review provides a structured reference for researchers, and establishes a foundation for the development of a generalized, unified EVCD modeling framework.

\bibliographystyle{IEEEtran}
\vspace{-5pt}
\bibliography{references}
\vspace{-3pt}

\end{document}